\date{}
\documentclass[showpacs,superscriptaddress]{revtex4}
\usepackage{epsf}
\usepackage{graphicx,amsmath,amssymb,amsfonts,bbm,latexsym,color,dcolumn,epsf}

\begin{document}
\title{Hertz potentials approach to the dynamical Casimir effect in cylindrical cavities of arbitrary section}

\author{Mart\'\i n Crocce}
\affiliation{Physics Department, New York University, 4 Washington Place, New York, New York 10003}

\author{Diego A. R. Dalvit}
\affiliation{Theoretical Division, MS B213 Los Alamos National Laboratory, Los Alamos, NM 87545}

\author{Fernando C. Lombardo}
\author{ Francisco D. Mazzitelli}
\affiliation{Departamento de F\'\i sica {\it Juan Jos\'e Giambiagi}, FCEyN UBA,
Facultad de Ciencias Exactas y Naturales, Ciudad Universitaria,
Pabell\' on I, 1428 Buenos Aires, Argentina}

\date{today}

\begin{abstract}
We study the creation of photons  in resonant cylindrical cavities
with time dependent length. The physical degrees of freedom of the
electromagnetic field are described using Hertz potentials. We
describe the general formalism  for cavities with arbitrary
section. Then we compute explicitly the number of TE and TM
motion-induced photons for cylindrical cavities with rectangular and circular
sections. We also discuss the creation of TEM photons in
non-simply connected cylindrical cavities.
\end{abstract}

\pacs{03.70.+k}

\maketitle

\newcommand{\beq}{\begin{equation}}
\newcommand{\eeq}{\end{equation}}
\newcommand{\dalam}{\nabla^2-\partial_t^2}


\section{Introduction}

In quantum field theory time-dependent boundary conditions or
time-dependent background fields may induce particle creation,
even when the initial state of a quantum field is the vacuum
\cite{birrel}. In the context of quantum electrodynamics,
uncharged mirrors in accelerated motion can in principle create
photons. This effect is referred to in the literature as the
dynamical Casimir effect, or motion-induced radiation
\cite{reviewdod}. In particular, when the length of a high Q
electromagnetic cavity oscillates with one of its resonant
frequencies, the number of photons inside the cavity accumulates
slowly and grows exponentially with time. Many authors have
considered this problem using different approximations: from toy
models of scalar fields in $1+1$ dimensions \cite{varios1+1} to
the more realistic case of scalar \cite{us,others} and
electromagnetic \cite{martin} fields in three dimensional rectangular cavities.
Arbitrary periodic motion of the boundary of an ideal cavity has
been studied in \cite{vcubo}. The relevance of finite temperature
effects  and losses have also been considered \cite{plunien}.

Unlike the static Casimir effect \cite{mostep}, that has been
measured with increasing precision in the last years \cite{exp},
an experimental verification  of the dynamical counterpart is
still lacking.  The main reason is that typical resonance
frequencies for microwave cavities  are of the order of GHz.  It
is very difficult (although not impossible) to make a mirror
oscillate at such frequencies. One possibility is to consider a
two dimensional array of nanoresonators coherently driven to
oscillate at very high frequencies \cite{onofrio}.

Several alternative proposals have been investigated in which the
physical properties of the medium inside the cavity change with
time, but keeping fixed the boundary of the cavity. For example,
it has been proposed \cite{lozo} that one could change effectively
the length of a cavity by irradiating with ultra-short laser
pulses a thin semiconductor film deposited on one of the walls of
the cavity (see also \cite{yablo}). Nonlinear optics may be used
to produce effective fast moving mirrors.

From an experimental point of view, the idea of changing the
effective length of a cavity by irradiating a semiconductor is
promising \cite{carugno}. A relevant possibility is to force
periodic oscillations of the conductivity of the semiconductor, as
this will enhance particle production for certain resonance
frequencies \cite{nos04}. The first estimations of the number of
photon created in resonant situations are within the limits of the
minimum signal that could be detected.  It is then of interest to
refine the calculations and to explore other geometries that could
be relevant from an experimental point of view. In order to
understand the effect of the geometry, in this paper we will
analyze the moving mirror case for cylindrical cavities of
arbitrary section (see Fig. 1). The physical degrees of freedom of
the electromagnetic field will be expressed in terms of the so
called scalar Hertz potentials. We will reobtain previous results
for rectangular cavities, and we will extend them to the case of
cylindrical cavities with circular section. We will also consider
the case of non simply connected cylindrical cavities, since in
this case one has additional Transverse Electromagnetic (TEM)
modes.

The paper is organized as follows. In Section II we introduce the
Hertz potentials and we express the boundary conditions for moving
mirrors in terms of them. We also show the equivalence of the
description with previous approaches based on dual Transverse
Electric (TE) and Magnetic (TM) vector potentials. In Section III
we quantize the theory, and we find the relation between the
number of TE and TM created photons with the number of  particles
associated with the scalar Hertz potentials. Section IV includes
explicit calculations for cavities with rectangular and circular
sections. The case of coaxial resonant cavities is considered in
Section V, where we quantize the TEM modes and show that they can
be described by a one dimensional scalar field . Section VI
contains a summary and a discussion of the results.

\begin{figure}[t]
\begin{center}
\includegraphics[width=0.4\textwidth]{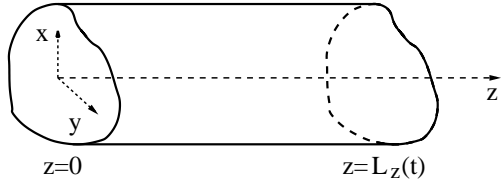}
\caption{Hollow, cylindrical cavity of arbitrary cross-sectional shape.}
\label{figure1}
\end{center}
\end{figure}

\section{Classical description}

\subsection{Vector and scalar Hertz potentials}

We consider in this section the representation of the physical
degrees of freedom of the  electromagnetic field in terms of Hertz
potentials, which allows an alternative picture to the standard
vector ${\bf A}$ and scalar potentials $\Phi$. In the Lorentz
gauge $\mu \epsilon \partial_t \Phi + \nabla \cdot  {\bf A} = 0$,
Maxwell equations for linear media read
\begin{eqnarray}
\mu\epsilon\frac{\partial^2\Phi}{\partial t^2}-\nabla^2
\Phi&=&\frac{1}{\epsilon}\rho-\frac{1}{\epsilon} \nabla\cdot{\bf
P}_0,
\label{eqPhi} \\
\mu\epsilon\frac{\partial^2{\bf A}}{\partial t^2}-\nabla^2
{\bf A}&=&\mu {\bf J}+\mu \frac{\partial
{\bf P}_0}{\partial t}+\nabla\times{\bf M}_0 , \label{eqA}
\end{eqnarray}
where we have included, in addition to the the induced electric
and magnetic polarizations (which are taken into account by the
constants  $\epsilon$ and $\mu$), permanent  polarizations and
magnetization ${\bf P}_0$ and ${\bf M}_0$. These are introduced
to  motivate the form of the Hertz potentials \cite{nisbet,rocio}.
We are using SI units with $\epsilon_0=1$, $\mu_0=1$ and
$c=1$.

Now we introduce two vector potentials, ${\bf \Pi}_e$ and ${\bf \Pi}_m$, by
expressing $\Phi$ and ${\bf A}$ in a symmetric form with respect
to the terms containing ${\bf P}_0$ and ${\bf M}_0$ in the r.h.s.
of Eqs. (\ref{eqPhi},\ref{eqA}),
\begin{eqnarray}
\Phi&=&-\frac{1}{\epsilon} \nabla\cdot{\bf \Pi}_e ,
\label{Phi}       \\
{\bf A}&=&\mu \frac{\partial {\bf \Pi}_e}{\partial t}+\nabla\times{\bf \Pi}_m . \label{A}
\end{eqnarray}
${\bf \Pi}_e$ and ${\bf \Pi}_m$ are known as the electric and magnetic  vector Hertz
potentials \cite{jackson,nisbet}. The equations satisfied
by the these potentials are simplified if one introduces
two functions, ${\bf Q}_e$ and ${\bf Q}_m$, known as stream
potentials, as follows
\begin{eqnarray}
\rho=-\nabla\cdot{\bf Q}_e  &;&
{\bf J}=\frac{\partial {\bf Q}_e}{\partial t} +
\frac{1}{\mu} \nabla \times {\bf Q}_m.
\label{streampotentials}
\end{eqnarray}
Replacing Eqs. (\ref{Phi}, \ref{A}, \ref{streampotentials}) into
Eqs. (\ref{eqPhi},\ref{eqA}) one finally arrives to
\begin{eqnarray}
(\mu \epsilon \, \partial_t^2 - \nabla^2) \,{\bf \Pi}_e &=&{\bf Q}_e + {\bf P}_0 , \label{waveeq1}\\
(\mu \epsilon \, \partial_t^2 - \nabla^2) {\bf \Pi}_m &=&{\bf Q}_m +
{\bf M}_0 .
\label{waveeqs}
\end{eqnarray}
There is a gauge freedom that must be suitably fixed in order to
obtain Eqs.(\ref{waveeq1}) and (\ref{waveeqs}) \cite{jackson}.
The electric field and magnetic induction are given in terms of
the vector Hertz  potentials by
\begin{eqnarray}
{\bf E} &=& \frac{1}{\epsilon} \nabla (\nabla\cdot {\bf \Pi}_e) - \mu \frac{\partial^2{\bf \Pi}_e}{\partial t^2}-\nabla\times\frac{\partial{\bf \Pi}_m}{\partial t}  ,  \label{electric} \\
{\bf B} &=& \mu \nabla\times\frac{\partial {\bf \Pi}_e}{\partial
t} + \nabla\times(\nabla\times{\bf \Pi}_m).
\label{magnetic}
\end{eqnarray}

In vacuum at points away from the sources, the fields can be
expressed in terms of only two scalar functions, the so-called
scalar Hertz potentials,  each satisfying an homogeneous wave
equation \cite{nisbet}. This result translates to the Hertz
formalism by writting

\begin{eqnarray}
{\bf \Pi}_e=\phi \,{\bf \hat e}_3  &;&  {\bf \Pi}_m=\psi \,{\bf \hat e}_3 ,
\label{whittaker}
\end{eqnarray}
where ${\bf \hat e}_3$ is a unit vector in a fixed direction. In this way $\psi$ gives rise to
TE fields with respect to ${\bf \hat e}_3$, whereas $\phi$ represents TM fields.
For our purposes ${\bf \hat e}_3$  will be taken along the longitudinal axis of the
cylindrical cavity. Moreover we will use either cartesian or cylindrical coordinates in this paper, with
${\bf  \hat e}_3={\bf \hat z}$ in both cases. The transverse directions $({\bf \hat e}_1,{\bf \hat e}_2)$ will stand for $({\bf \hat x},{\bf \hat y})$
in the first case and $({\bf \hat \rho},{\bf \hat \phi})$ in the second.

The potentials and fields can be written in terms of the two
scalar Hertz potentials replacing Eq. (\ref{whittaker}) into
Eqs. (\ref{Phi},\ref{A},\ref{electric},\ref{magnetic}). The result
is
\begin{eqnarray}
\Phi&=&-\partial_z\phi , \\
{\bf A}&=&\partial_2\psi\,{\bf \hat e}_1-\partial_1\psi\,{\bf \hat e}_2+\partial_t\phi\,{\bf \hat z} , \label{Amu}\\
{\bf E}&=&(\partial_1\partial_z\phi-\partial_2\partial_t\psi)\,{\bf \hat e}_1+(\partial_2\partial_z\phi+\partial_1\partial_t\psi)\,{\bf \hat e}_2-(\nabla^2_{\perp}\phi)\,{\bf \hat z}  ,  \label{E}  \\
{\bf B}&=&(\partial_2\partial_t\phi+\partial_1\partial_z\psi)\,{\bf \hat e}_1+(-\partial_1\partial_t\phi+\partial_2\partial_z\psi)\,{\bf \hat e}_2-(\nabla^2_{\perp}\psi)\,{\bf \hat z} , \label{B}
\end{eqnarray}
where the transverse Laplacian is defined as
$\nabla^2_{\perp}\equiv\nabla^2 - \frac {\partial^2} {\partial
z^2}$. In cartesian coordinates
$(\partial_1,\partial_2)=(\frac{\partial}{\partial
x},\frac{\partial}{\partial y})$ and
$\nabla^2_{\perp}=\frac{\partial^2}{\partial
x^2}+\frac{\partial^2}{\partial y^2}$. In cylindrical coordinates
$(\partial_1,\partial_2)=(\frac{\partial}{\partial
\rho},\frac{1}{\rho}\frac{\partial}{\partial \phi})$ and
$\nabla^2_{\perp}=\frac{1}{\rho} \frac{\partial}{\partial
\rho}\left(\frac{1}{\rho}\frac{\partial}{\partial
\rho}\right)+\frac{1}{\rho^2}\frac{\partial}{\partial \phi^2}$.

In previous works \cite{paulo,martin} the electromagnetic degrees
of freedom have been described in terms of two vector potentials
${\bf A}_{\rm TE}$ and ${\bf A}_{\rm TM}$ with null divergence
and ${\bf z}$-component. The TE electric and magnetic fields are
given by
\begin{eqnarray}
 {\bf E}_{\rm TE}=-\dot{\bf A}_{\rm TE} &;& {\bf B}_{\rm TE}=\nabla\times{\bf A}_{\rm TE},
\label{dualE}
\end{eqnarray}
while the TM fields are given by the dual relations
\begin{eqnarray}
{\bf B}_{\rm TM}=\dot{\bf A}_{\rm TM} &;& {\bf E}_{\rm TM}=\nabla\times{\bf A}_{\rm TM}
\label{dualM} .
\end{eqnarray}
Comparing Eqs. (\ref{dualE}) and (\ref{dualM}) with the expressions
for the electromagnetic fields in terms of the Hertz potentials Eqs. (\ref{electric},\ref{magnetic}) and Eq. (\ref{whittaker}) we
obtain
\begin{eqnarray}
{\bf A}_{\rm TE}&=&\nabla\times{\bf \Pi}_m ={\bf\hat z}\times\nabla\psi , \nonumber\\
{\bf A}_{\rm TM}&=&\nabla\times{\bf \Pi}_e =  {\bf\hat
z}\times\nabla\phi ,\label{relations}
\end{eqnarray}
so both approaches are equivalent.

The description in terms of independent TE and TM fields is
possible due to the particular geometries we are considering.
Indeed, using the above definitions and boundary conditions (see
below) it is easy to check that no mixed terms appear in Maxwell's
Lagrangian and Hamiltonian.

\subsection{Boundary conditions for a perfect conducting cavity}
\label{standardsolution}

In the static case, the boundary conditions for the fields over
the walls of the cavity are ${\bf E}_t =0$ and ${\bf B}_{n}=0$.
These translate into conditions for the two scalar potentials
through Eqs. (\ref{E},\ref{B}). The scalar Hertz potential $\psi$
satisfies Dirichlet boundary conditions on the longitudinal
boundary ($z=0,L_z$) and Neumann boundary conditions on the
transverse boundaries,
\begin{eqnarray}
\psi |_{z=0,L_z}=0   &;&
\frac{\partial \psi}{\partial n}\left|_{{\rm trans}}\right.=0 .
\label{bcpsi}
\end{eqnarray}
On the other hand, the scalar Hertz potential $\phi$ satisfies
Neumann boundary conditions on the longitudinal boundary
($z=0,L_z$) and Dirichlet boundary conditions on the transverse
boundaries,
\begin{eqnarray}
\frac{\partial \phi}{\partial z}\left|_{z=0,L_z}\right.=0  &;&  \phi |_{{\rm trans}}=0 .
\label{bcphi}
\end{eqnarray}

Let us now consider the boundary conditions in the case when one
of the surfaces is moving, say $z=L_z(t)$. The boundary conditions on a
moving interface between two mediums containing a surface charge
density $\sigma$ and a surface current ${\bf K}$ are \cite{namias}
\begin{eqnarray}
({\bf D}_{\rm II}-{\bf D}_{\rm I})\cdot {\bf n}&=&\sigma ,    \nonumber     \\
({\bf B}_{\rm II}-{\bf B}_{\rm I})\cdot {\bf n}&=& 0 , \nonumber       \\
\left[{\bf n}\times({\bf H}_{\rm II}-{\bf H}_{\rm I})+({\bf v}\cdot{\bf n})({\bf D}_{\rm II}-{\bf D}_{\rm I})\right]\cdot {\bf t}&=&{\bf K}\cdot {\bf t}  , \nonumber              \\
\left[{\bf n}\times({\bf E}_{\rm II}-{\bf E}_{\rm I})-({\bf
v}\cdot{\bf n})({\bf B}_{\rm II}-{\bf B}_{\rm I})\right]\cdot
{\bf t}&=&0,      \label{bc}
\end{eqnarray}
where ${\bf n}$ denotes the normal to the interface going from
medium I to medium II, and ${\bf t}$ is any unit vector
tangential to the surface. These conditions can be derived by
performing a Lorentz transformation to the reference system where
the surface is instantaneously at rest \cite{paulo}, or by
consistency of the Maxwell equations in the laboratory frame
\cite{namias}.

We assume the moving wall to be a perfect conductor.
Therefore the fields vanish exactly in region II and the
boundary conditions in Eq. (\ref{bc}) become
\begin{eqnarray}
{\bf B}\cdot{\bf\hat{z}} =0 &;&
( {\bf E} \times {\bf \hat{z}} + v  {\bf B} ) \cdot {\bf t} =0,
\end{eqnarray}
where $v=\dot{L}_z$. In terms of the two scalar Hertz potentials
we obtain
\begin{eqnarray}
\psi(z=L_z(t))=0  &;&
(\partial_z + v\,\partial_0)\phi(z=L_z(t))=0,
\end{eqnarray}
that modify the boundary conditions in Eqs. (\ref{bcpsi}) and
(\ref{bcphi}) on the moving longitudinal boundary ($z=L_z(t)$).

\section{Quantum description}

In this section we will quantize the electromagnetic field using
the scalar Hertz potentials formalism. We will also obtain the
relation between the Bogoliubov coefficients relating IN and OUT
basis of the scalar field with those of the electromagnetic
field. We will assume that the cavity is at rest for $t<0$, and
that the wall placed at $z=L_z$ begins to move with a prescribed
trajectory $L_z(t)$. We will mainly concentrate on harmonic
motions of the type
\begin{equation}
L_z(t) = L_0 \left[1 + \epsilon \sin\Omega t + \epsilon
f(t)\right],
\end{equation}
where $f(t)$ is some decaying function that allows to meet the
continuity conditions at $t=0$. The motion ends at $t=T$.

\subsection{Transverse electric scalar field}

Let us start with the TE scalar Hertz potential $\psi$. We expand the
field as
\begin{equation}
\psi({\bf x},t) =\sum_{{\bf k}}\,C_{\bf k} a_{\bf k}^{\rm IN} u^{\rm IN}_{{\bf k},{\rm TE}}({\bf x},t) + h.c. ,
\end{equation}
where $u^{\rm IN}_{{\bf k},{\rm TE}}({\bf x},t)$ are the
solutions of the Klein Gordon equation which have positive
frequency in the IN region ($t<0$), $a_{\bf k}^{\rm IN}$ are the
corresponding annihilation bosonic operators, and $C_{\bf k}$ are
normalization constants. The summation index is ${\bf k}=({\bf
k}_{\perp},k_z=n_z \pi/L_z)$. In the IN region the explicit form
of the solution is
\begin{equation}
u^{\rm IN}_{{\bf k},{\rm TE}}({\bf x},t)=
\frac{e^{-i\omega_{\bf k} t}}{\sqrt{2\omega_{\bf k}}}
\sqrt{\frac{2}{L_z}} \sin (k_z z) v_ {{\bf k}_{\perp}}({\bf x}_{\perp}) ,
\label{IN}
\end{equation}
with $\nabla^2_{\perp}v_ {{\bf k}_{\perp}}=-{\bf k}^2_{\perp}v_ {{\bf k}_{\perp}}$
and $w_{\bf k}=\vert {\bf k}\vert$. The set of functions $v_ {{\bf k}_{\perp}}$ satisfy
Neumann boundary conditions on the lateral surface and can be
assumed to be real and orthonormal on the plane ${\bf x}_{\perp}$.

In order to fix the normalization constants $C_{\bf k}$ and the
commutation relations between creation and annihilation operators,
we compute the Hamiltonian of the electromagnetic field. We
find
\begin{eqnarray}
H &=& \frac{1}{8\pi}\int d^3x~({\bf E}^2+{\bf B}^2)= \frac{1}{8\pi}\int d^3x
(-\dot\psi\nabla^2_{\perp}\dot\psi+\nabla^2\psi\nabla^2_{\perp}\psi)
\nonumber\\
&=&\sum_{{\bf k}}\,\frac{\vert C_{\bf k}\vert^2}{8\pi}{\bf k}^2_{\perp}\omega_{\bf
k}(a_{\bf k}^{\rm IN} (a_{\bf k}^{\rm IN})^{\dagger} + h.c.)
\end{eqnarray}
Therefore, with the choice $\vert C_{\bf k}\vert = \sqrt{8\pi}/\vert{\bf
k}_{\perp}\vert$, the operators $a_{\bf k}^{\rm IN}$ and
$(a_{\bf k}^{\rm IN})^{\dagger}$ satisfy the usual commutation relations \cite{rocio}.

For $0<t<T$ we expand the IN basis in an instantaneous basis \beq
u^{\rm IN}_{{\bf k},{\rm TE}} = \sum_{\bf p} Q_{{\bf p},{\rm
TE}}^{(\bf k)}(t) \ \sqrt{\frac{2}{L_z(t)}} \sin (p_z(t) z)   v_
{{\bf p}_{\perp}}({\bf x}_{\perp}), \eeq where $p_z(t)=p_z
\pi/L_z(t)$. The initial conditions are
\begin{eqnarray}
Q_{{\bf p},{\rm TE}}^{(\bf k)}(0) = \frac{1}{\sqrt{2\omega_{\bf k}}}
\delta_{{\bf p}{\bf k}}  &;& {\dot Q}_{{\bf p},{\rm TE}}^{(\bf k)}(0)
= - i \sqrt{\frac{\omega_{\bf k}}{2}}\delta_{{\bf p}{\bf k}} .
\end{eqnarray}
This expansion must be a solution of the wave equation. Using the
fact that both $\sin(p_z(t) z)$ and $ v_ {{\bf p}_{\perp}}({\bf
x}_{\perp})$ form a complete and orthonormal set, and the fact
that they only depend on $t$ through $L_z(t)$, we obtain a set of
coupled equations for the temporal coefficients $Q_{{\bf p},{\rm
TE}}^{(\bf k)}(t)$ \cite{martin}
\begin{eqnarray}
\ddot{Q}_{{\bf p},{\rm TE}}^{(\bf k)} + \omega_{\bf p}^2(t) Q_{{\bf p},{\rm TE}}^{(\bf k)}
&=& 2 \lambda (t) \sum_{\bf j} g_{\bf p\bf j}~
\dot{Q}_{{\bf j},{\rm TE}}^{(\bf k)} + \dot\lambda (t) \sum_{\bf j} g_{{\bf p}{\bf j}}
Q_{{\bf j},{\rm TE}}^{(\bf k)} \nonumber \\
&+& O(\epsilon^2) ,
\label{Qeq}
\end{eqnarray}
where
\begin{eqnarray}
\omega_{\bf p}(t) = \sqrt{ {\bf p}_\perp^2 + \left(\frac{p_z \pi}{L_z(t)}\right)^2}  &;&
\lambda (t) = \frac{\dot{L}_z(t)}{L_z(t)}.
\end{eqnarray}
The coupling coefficients are given by
\begin{equation}
g_{\mathbf{pj}}=-g_{\mathbf{jp}}= \left\{
\begin{array}{ll}
(-1)^{p_{z}+j_{z}}\frac{2p_{z}j_{z}}{j_{z}^{2}-p_{z}^{2}} \,
\delta_{{\bf p}_{\perp}, {\bf j}_{\perp}} & \mbox{if $p_{z}\neq j_{z}$} \\
0 & \mbox{if $p_{z}=j_{z}$}  .
\end{array}
\right.
\end{equation}

For $t>T$ (OUT region), the time dependent coefficients $Q_{{\bf p},{\rm TE}}^{\bf k}$
become
\begin{equation}
 Q_{{\bf p},{\rm TE}}^{\bf k} = A_{{\bf p},{\rm TE}}^{\bf k} {e^{-i\omega_{\bf k} t}
 \over\sqrt{2\omega_{\bf
k}}} + B_{{\bf p},{\rm TE}}^{\bf k} {e^{i\omega_{\bf k} t}\over\sqrt{2\omega_{\bf k}}} .
\end{equation}
We introduce the OUT basis  $u^{\rm OUT}_{{\bf k},{\rm TE}}({\bf x},t)$ as the set
of solutions  of the Klein-Gordon equation that are of the form
given in the r.h.s of Eq. (\ref{IN}) in the OUT region. The IN and
OUT basis are then related by the Bogoliubov transformation
\begin{equation}
u^{\rm IN}_{{\bf k},{\rm TE}}=\sum_{\bf p} B^{\bf k}_{{\bf
p},{\rm TE}} (u^{\rm OUT}_{{\bf p},{\rm TE}})^* + A^{\bf k}_{{\bf
p},{\rm TE}} u^{\rm OUT }_{{\bf p},{\rm TE}} . \label{bogo}
\end{equation}
Using this relation it is easy to show that the number of OUT photons with TE polarization  and
wavevector ${\bf k}$ is given by
\begin{equation}
\langle N_{{\bf k},{\rm TE}} \rangle =
\langle 0_{\rm IN} | (a_{\bf k}^{\rm OUT})^{\dagger} a_{\bf k}^{\rm OUT} | 0_{\rm IN} \rangle =
{\bf k}^2_{\perp} \sum_{\bf p}{\vert B^{\bf p}_{{\bf k},{\rm TE}} \vert^2\over {\bf
p}^2_{\perp}} .
\label{npart}
\end{equation}

\subsection{Transverse magnetic scalar field}

The quantization of the TM scalar Hertz potential $\phi$ is
analogous to that of $\psi$. The explicit form for the IN basis is now given by
\begin{equation}
u^{\rm IN}_{{\bf k},{\rm TM}}({\bf x},t)=
\frac{e^{-i\omega_{\bf k}t}}{\sqrt{2\omega_{\bf k}}}
\sqrt{\frac{2}{L_z }} \cos (k_z z) r_ {{\bf k}_{\perp}}({\bf x}_{\perp}) ,
\label{INTM}
\end{equation}
with $\nabla^2_{\perp}r_ {{\bf k}_{\perp}}=-{\bf k}^2_{\perp}r_ {{\bf
k}_{\perp}}$, the functions $r_ {{\bf k}_{\perp}}$ satisfying
Dirichlet boundary conditions on the lateral surface. Following
\cite{martin}, we introduce the instantaneous basis
\begin{equation}
u^{\rm IN}_{{\bf k},{\rm TM}} =
\sum_{\bf p} ( Q^{({\bf k})}_{{\bf p},{\rm TM}}(t) + \dot{Q}^{({\bf k})}_{{\bf p},{\rm TM}}(t) g(z,t) )
\sqrt{\frac{2}{L_z(t)}} \cos(p_z(t) z) r_ {{\bf k}_{\perp}}({\bf x}_{\perp}).
\end{equation}
The initial conditions for $Q^{({\bf k})}_{{\bf p},{\rm TM}}$ are
the same as those for $Q^{({\bf k})}_{{\bf p},{\rm TE}}$. The
function $g(z,t)$ can be expressed as $g(z,t)=\dot{L}_z(t)
L_z(t)  \xi(z/L_z(t))$, where $\xi(z)$ is a solution to the
conditions $\xi(0)=\xi(1)=0$, $\partial_z \xi(0)=0$, and
$\partial_z \xi(1) = -1$ \cite{martin}. There are many solutions
to these conditions, implying a freedom for selecting the
instantaneous basis. However, it can be proved that physical
quantities, such as the mean number of created TM photons or the
energy density inside the cavity are independent of the
particular choice of $g(z,t)$ \cite{martin}. The equation of
motion for $Q^{({\bf k})}_{{\bf p},{\rm TM}}$ is similar to Eq.
(\ref{Qeq}) for  $Q^{({\bf k})}_{{\bf p},{\rm TE}}$, namely
\begin{eqnarray}
\ddot{Q}^{({\bf k})}_{{\bf p},{\rm TM}} + \omega^2_{\bf k}(t) Q^{({\bf k})}_{{\bf p},{\rm TM}} &=&
-2 \lambda(t) \sum_{\bf j} h_{\bf j p} \dot{Q}^{({\bf k})}_{{\bf p},{\rm TM}}
- \dot{\lambda}(t) \sum_{\bf j} h_{\bf j p} Q^{({\bf k})}_{{\bf p},{\rm TM}} \nonumber \\
&&
-2 \dot{\lambda}(t) L_z^2(t) \sum_{\bf j} s_{\bf j p} \ddot{Q}^{({\bf k})}_{{\bf p},{\rm TM}}
- \sum_{\bf j} \dot{Q}^{({\bf k})}_{{\bf p},{\rm TM}}
[ s_{\bf j p} \ddot{\lambda}(t) L_z^2(t) - \lambda(t) \eta_{\bf j p} ] \nonumber \\
&& - \lambda(t) L_z^2(t) \sum_{\bf j} s_{\bf j p} \partial_t^3
{Q}^{({\bf k})}_{{\bf p},{\rm TM}} + O(\epsilon^2)  ,
\label{QTM}
\end{eqnarray}
where the coefficients $s_{\bf j p}$, $\eta_{\bf j p}$ and $h_{\bf j p}$ are given by
\begin{eqnarray}
s_{\bf j p} &=& \int_0^{L_z(t)} dz \xi(z) \chi_{\bf j} \chi_{\bf p} , \nonumber \\
\eta_{\bf j p} &=& L_z^2(t) \int_0^{L_z(t)} dz [ ( \xi^{''}(z) - \omega_{\bf j}^2 \xi(z)) \chi_{\bf j} \chi_{\bf k}
+ 2 \xi^{'}(z)  \chi^{'}_{\bf j} \chi_{\bf k} ] ,\nonumber \\
h_{\bf{j p}} &=& \left\{
\begin{array}{ll}
(-1)^{p_{z}+j_{z}}\frac{2j_{z}^2}{p_{z}^{2}-j_{z}^{2}} \,
\delta_{{\bf p}_{\perp}, {\bf j}_{\perp}} & \mbox{if $p_{z}\neq j_{z}$} \\
-   \delta_{{\bf p}_{\perp}, {\bf j}_{\perp}} & \mbox{if $p_{z}=j_{z}$}  .
\end{array}
\right.\nonumber
\end{eqnarray}
Here $'$ denotes derivative with respect to $z$. The functions
$\chi_{\bf j} \equiv \sqrt{2/L_z(t)} \sin(n_z \pi z/L_z(t))$ are
normalized in the interval $[0,L_z(t)]$. Note that the above
coefficients are independent of the particular form of the spatial
modes $r_{{\bf j}_{\perp}}({\bf x}_{\perp})$ for the transversal
section.

The equation for the number of TM photons in the OUT region is
similar to Eq. (\ref{npart}), with $B^{\bf p}_{{\bf k},{\rm TE}}$
replaced by $B^{\bf p}_{{\bf k},{\rm TM}}$. Eq. (\ref{npart}) and
its TM counterpart are very useful because they relate the number
of motion-induced photons with the Bogoliubov transformation of
two scalar fields. Therefore, the analysis is simplified since it
does not involve any reference to the polarization of the
electromagnetic field. Moreover, the information about the
transversal section of the cavity only enters through the spectrum
$\omega_{\bf k}$.

\section{Applications}

In this section we will compute the number of photons created in
resonant situations,  for both cylindrical cavities with
rectangular and circular sections. The equations of motion for
the coefficients $Q_{{\bf p},{\rm TE}}^{({\bf k})}$ of the scalar
Hertz potential $\Psi$ (Eq. (\ref{Qeq})) and the similar one for
the scalar Hertz potential $\phi$ (Eq. (\ref{QTM})) describe a
set of coupled harmonic oscillators with periodic frequencies and
couplings. They are of the same form as the equations that
describe the modes of a scalar field in a three dimensional
cavity with an oscillating boundary, and can be solved  using
multiple scale analysis (MSA). For a detailed description of the
method see, for example, \cite{us}.

In the "parametric resonant case", in which the external frequency
$\Omega$ is twice the frequency of an unperturbed mode ${\bf k}$
($\Omega=2 \omega_{\bf k}$), the equations Eq. (\ref{Qeq}) and Eq.
(\ref{QTM}) lead to a resonant behavior of the solutions.
Moreover, there is intermode coupling between modes ${\bf j}$ and
${\bf k}$ if any of the resonant coupling conditions $\Omega = |
\omega_{\bf k} \pm \omega_{\bf j} |$ is satisfied. This is the
case, for example,  for 1D geometries and cubic cavities. Except
for special geometries,  in general the resonant coupling
conditions are not met: different ${\bf k}$  modes will not be
coupled during the dynamics, and Eq. (\ref{Qeq}) and Eq.
(\ref{QTM}) reduce to the Mathieu equation for a single mode. In
consequence, the number of motion-induced photons in that given
mode will grow exponentially. The growth rate is different for TE
and TM modes. This is due to the fact that, while the r.h.s. of
Eq. (\ref{Qeq}) vanishes for ${\bf j}={\bf p}$, this is not the
case in Eq. (\ref{QTM}). The result is
\begin{eqnarray}
\langle N_{{\bf k},{\rm TE}}(t) \rangle &=& \sinh^2(\lambda_{{\bf k},{\rm TE}} \epsilon t)  , \label{LTE} \\
\langle N_{{\bf k},{\rm TM}}(t) \rangle &=& \sinh^2(\lambda_{{\bf k},{\rm TM}} \epsilon t)  \label{LTM} ,
\end{eqnarray}
where $\lambda_{{\bf k},{\rm TE}} = k_z^2 / 2 \omega_{\bf k}$ and
$\lambda_{{\bf k},{\rm TM}} = (2 \omega^2_{\bf k} - k_z^2)/ 2
\omega_{\bf k}$. Note that when both polarizations are present,
the rate of growth for TM photons is larger than for TE photons,
i.e., $\lambda_{{\bf k},{\rm TM}} > \lambda_{{\bf k},{\rm TE}}$.
As already mentioned, these expressions are independent of the
mode functions in the transverse direction, $v_{{\bf
k}_{\perp}}({\bf x}_{\perp})$ or $r_{{\bf k}_{\perp}}({\bf
x}_{\perp})$. The dependence on the sectional geometry enters
only through the spectrum $\omega_{\bf k}$. The above expressions
for the growth rates Eqs. (\ref{LTE}, \ref{LTM}) are valid for
uncoupled modes. When there is mode coupling, the expressions for
the growth rates are different, as explained in \cite{us,martin}.

\subsection{Cavities with rectangular section}

This problem has been analyzed previously in \cite{martin}. The
transverse mode functions $v_{{\bf k}_{\perp}}({\bf x}_{\perp})$
associated with the scalar Hertz potential $\psi$ are
\begin{equation}
v_{n_x,n_y}({\bf x}_{\perp}) = \frac{2}{\sqrt{L_x L_y}}  \cos \left( \frac{n_x \pi x}{L_x} \right)
\cos \left( \frac{n_y \pi y}{L_y} \right),
\end{equation}
where $n_x$
and $n_y$ are non-negative integers that cannot be simultaneously
zero. The spectrum is ($n_z\ge 1$),
\begin{equation}
\omega_{n_x,n_y,n_z} = \sqrt{(n_x \pi/L_x)^2 + (n_y \pi/L_y)^2 + (n_z \pi/L_z)^2}.
\label{rectangularespectrum}
\end{equation}
The transverse mode functions $r_{{\bf k}_{\perp}}({\bf x}_{\perp})$ associated
with the scalar Hertz potential $\phi$ are
\begin{equation}
r_{m_x,m_y}({\bf x}_{\perp}) = \frac{2}{\sqrt{L_x L_y}} \sin \left( \frac{m_x \pi x}{L_x} \right)
\sin \left( \frac{m_y \pi y}{L_y} \right),
\end{equation}
where $m_x$ and $m_y$ are integers, such that $m_x, m_y \ge 1$. The spectrum is given by
Eq. (\ref{rectangularespectrum}) with $n_z\ge 0$.

As an example, we consider the parametric resonant case $\Omega=2
\omega_{\bf k}$, and show the results for the number of photons
created in a cubic cavity of size $L$. Let us denote the modes by
$(n_x,n_y,n_z)$. For the TE case, the fundamental mode is doubly
degenerate ($(1,0,1)$ and $(0,1,1)$) and uncoupled to other higher
modes. The number of TE photons in these modes grows exponentially
as $\exp(\pi \epsilon t / \sqrt{2} L)$. The fundamental mode for
the TM fields corresponds to $(1,1,0)$, and is coupled to the mode
$(1,1,4)$.  The number of TM photons in these modes grows as
$\exp(4.4 \epsilon t / L)$ \cite{martin}. The lowest possible
excitable frequency of the field equals the fundamental frequency
of both the TE and TM spectrums. However the exponential growth of
motion-induced TM  photons with fundamental energy in greater than
that of TE photons.

\subsection{Cavities with circular section}

Let us now introduce cylindrical coordinates $(\rho,\phi,z)$ to
consider a cylinder with circular section of radius $\rho=R$. The
transverse mode functions $v_{{\bf k}_{\perp}}({\bf x}_{\perp})$
related to TE fields are

\beq v_{nm}({\bf x}_{\perp})=\frac{1}{\sqrt{\pi}}
\frac{1}{R J_n(y_{nm}) \sqrt{1-n^2/y_{nm}^2}}
J_n \left( y_{nm}\frac{\rho}{R} \right) \ \
e^{i n \phi} ,
\eeq where $J_n$ denotes the Bessel function of {\it n}th order,
and $y_{nm}$ is the {\it m}th positive root of the equation
$J'_n(y)=0$.
The eigenfrequencies are given by ($n_z \ge 1$)
\beq
\omega_{n,m,n_z}=\sqrt{\left(\frac{y_{nm}}{R}\right)^2+\left(\frac{n_z
\pi}{L_z}\right)^2}. \label{espectrocilindrico} \eeq

The solution for the mode functions  $r_{{\bf k}_{\perp}}({\bf
x}_{\perp})$ associated with TM fields are
\beq
r_{nm}({\bf x}_{\perp})=\frac{1}{\sqrt{\pi}} \frac{1}{R J_{n+1}(x_{nm})} \ \
J_n \left( x_{nm} \frac{\rho}{R} \right) \ \ e^{i n \phi} ,
\eeq
where $x_{nm}$ is the {\it m}th root of the equation $J_n(x)=0$. The spectrum is given by
Eq. (\ref{espectrocilindrico}) with $y_{nm}$ replaced by $x_{nm}$ and $n_z \ge 0$.

Denoting the modes by $(n,m,n_z)$, the lowest TE mode is
$(1,1,1)$ and has a frequency
$\omega_{111}=(1.841/R)\sqrt{1+ 2.912 (R/L_z)^2}$.  This mode is uncoupled to any other
modes, and according to Eq. (\ref{LTE}) the number of photons in this mode grows
exponentially in time as $\exp{(\pi \epsilon t / \sqrt{1+ 0.343 (L_z/R)^2} L_z)}$ when parametrically
excited. It is interesting to point out that for this mode it is
possible to tune the resonance by changing the relation of the
radius and the height of the cavity.
The lowest TM mode $(0,1,0)$ is also uncoupled and has a frequency
$\omega_{010}=2.405/R$. From Eq. (\ref{LTM}) we find the
parametric growth to be $\exp{(4.81 \epsilon  t/R)}$. This mode cannot be
resonated by simple tuning since its frequency depends solely on the radius of the cylinder.
 For $L_z$ large enough ($L_z > 2.03 R$), the resonance frequency $\omega_{111}$ of the lowest
TE mode is smaller than that for the lowest TM mode. Then the $(1,1,1)$ TE mode is the fundamental
oscillation of the cavity.

\subsection{Numerical estimations}

The number of generated photons computed in the previous Sections
is in general proportional to $\exp[2\lambda\epsilon t]$, where
$\lambda$ depends on the geometry and the particular mode
considered (see for example Eqs. (38) and (39), that are valid for
uncoupled modes). The amount of created photons is limited by the
$Q$ factor of the cavity. If the mirror oscillates during a time
$t_{max}\simeq Q/\omega$ (with $\omega$ the frequency of the
mode), the maximum number of particles is
$\exp[\frac{2\lambda}{\omega}\epsilon Q]$. In Table I we show the
value of $\frac{2\lambda}{\omega}$ for the lowest modes in
cavities with rectangular and circular sections. The maximal
dimensionless amplitude for the mechanical oscillation of the
mirror is of order $\epsilon_{max}\simeq 10^{-8}$ \cite{others}.
Therefore, a very high $Q$ factor is needed to produce a large
number of photons.

\vspace{.3cm}
\begin{table}[h!]
\begin{tabular}{|p{0.8in}|p{0.8in}|p{0.4in}|}
\hline
\multicolumn{1}{|c}{Cavity} & \multicolumn{1}{|c|}{Mode}
& \multicolumn{1}{c|}{$2\lambda/\omega$} \\
\hline
Cubic & TE (1,0,1) & \hspace{.3cm}0.5 \\
Cubic & TE (0,1,1) & \hspace{.3cm}0.5 \\
Cubic & TM (1,1,0) & \hspace{.3cm}1.0\\
Cubic & TM (1,1,4) & \hspace{.3cm}0.3\\
Cylindrical & TM (0,1,0) & \hspace{.3cm}2.0 \\
Cylindrical & TE (1,1,1) & \hspace{.3cm}0.03 \\
\hline
\end{tabular}
\caption{Values of $\frac{2\lambda}{\omega}$ for different
cavities. In the cubic case, the values are independent of the
size $L$. For cylindrical cavities of length $L$ and circular
section of radius $R$, we are assuming $L/R=10$ }\label{table1}
\end{table}

The constraint over $Q$ can be relaxed if, instead of considering
moving mirrors, one considers a cavity containing a thin
semiconductor film. The effective length of the cavity can be
changed by irradiating the semiconductor
\cite{lozo,yablo,carugno,nos04}. In this situation, the maximum
number of created photons  is of order $\exp[a\tilde\epsilon Q]$
where $a=O(1)$ and $\tilde\epsilon$ depends on the properties of
the semiconductor and on the geometry of the cavity. For
reasonable values of the parameters one can have values of
$\tilde\epsilon$ as large as $\simeq 10^{-2}$ \cite{nos04}. In
this case, a large number of photons can be produced, even if the
$Q$ factor is not so high. For example, for conservative values
$Q=10^6$, $\tilde\epsilon =10^{-4}$, $a=1$ the number of created
photons is of order $10^{43}$.

\section{Non-simply connected cavities: transverse electromagnetic modes}

When the cylindrical cavity is non-simply connected, in addition to the TE and
TM modes one should also consider the TEM modes, for which both
the electric and magnetic fields have vanishing $z$ components.
The treatment of the TE and TM modes in these cavities is similar
to the case of hollow cylinders. However, to describe the TEM
modes it is necessary to introduce an additional scalar field
$\varphi(z,t)$. Indeed, working with the usual vector potential
${\bf A}$, the TEM solutions are of the form
\begin{eqnarray}
{\bf A}({\bf x}_{\perp},z,t)&=&{\bf A}_{\perp}({\bf x}_{\perp})\varphi(z,t) , \\
{\bf E}&=& - (\partial_t \varphi) \; {\bf A}_{\perp} ,          \\
{\bf B}&=& (\partial_z \varphi) \; {\bf \hat z} \times {\bf A}_{\perp} .
\end{eqnarray}

The transverse vector potential has vanishing rotor and
divergence, and zero tangential component on the transverse surfaces. Therefore, ${\bf A}_{\perp}$ is a solution of an {\it electrostatic} problem in the two transverse dimensions (in hollow cylindrical cavities the transverse potential vanishes and
TEM  modes do not exist). The scalar field $\varphi$ satisfies Dirichlet boundary conditions on the longitudinal boundaries
$z=0$ and $z=L_z(t)$, and the longitudinal wave
equation $(\partial_t^2-\partial_z^2)\varphi = 0$. For a static
cavity, the eigenfrequencies of the TEM modes are $w_n= n \pi / L_z$.
Note that this is an equidistant spectrum.
In the particular case of a resonant cavity formed with two
concentric  cylinders, the transverse vector potential is given by
${\bf A}_{\perp} = {\hat \rho}/ \rho$. However, it is
important to stress that the description in this section is valid for a non
simply connected cavity of arbitrary section.

In order to quantize these TEM modes, we proceed as in Section
III. The Hamiltonian associated with TEM modes is \beq H^{TEM} =
{1\over 8 \pi} \int d^2x_{\perp}dz ~({\bf E}^2+{\bf B}^2)= {1\over
8 \pi} \left (\int d^2x_{\perp} \vert {\bf A}_{\perp}\vert^2\right
)\int dz [(\partial_t\varphi )^2 + (\partial_z\varphi)^2] . \eeq
The above equation shows that the quantization of TEM modes is
equivalent to the quantization of a scalar field in $1+1$
dimensions with Dirichlet boundary conditions at $z=0$ and
$z=L_z(t)$. This problem has been previously studied as a toy
model for the dynamical Casimir effect \cite{varios1+1}. It is
interesting that this toy model describes TEM waves in non-simply
connected cavities.

Due to the conformal symmetry, in $1+1$ dimensions it is  possible
to write down an explicit expression for the modes in terms of a
single function satisfying the so called Moore equation
\cite{moore}. This equation can be solved, for example,  using a
renormalization group improvement of the perturbative solution
\cite{rg}. We will not repeat the analysis here, but just quote
the main results. As the eigenfrequencies are equidistant, there
is intermode coupling and the spectrum of created photons is
completely different from the $3+1$ case. If the external
frequency is $\Omega=q\pi/L_z$ with $q$ an integer, $q \geq 2$,
photons are created resonantly in all modes with $n=q+2j$, with
$j$ being a non-negative integer. The number of photons in each
mode does not grow exponentially, but the total energy inside the
cavity does.

Using the conformal symmetry  in $1+1$ dimensions, it is possible
to compute not only the total energy inside the cavity but also
the mean value of the energy density
\beq
\langle T_{00}^{\rm TEM}({\bf x},t) \rangle =
{1\over 8\pi}\vert {\bf A}_{\perp}\vert^2
\langle [(\partial_t\varphi )^2 + (\partial_z\varphi)^2] \rangle \equiv
{1\over 4\pi}\vert {\bf A}_{\perp}({\bf x}_{\perp})\vert^2
\langle T_{00}(z,t) \rangle .
\eeq
It can be shown that the one-dimensional energy density $\langle T_{00}(z,t) \rangle$
grows exponentially in the form of $q$ travelling wave packets
which become narrower and higher as time increases. As an example,
in Fig. 2 we show the energy density profile as  a function of $z$
for a fixed $t$ and for the case $q=4$. As time evolves, the peaks
move back and forth bouncing against the caps of the cavity. The
height of the peaks increases as
 $e^{\frac{2 \pi q \epsilon t}{L_z}}$ and their width decreases as
 $e^{- \frac{\pi q \epsilon t}{L_z}}$, so that the total area
beneath each peak, and hence the total energy inside the cavity,
grows as $e^{\frac{\pi q \epsilon t}{L_z}}$. In Fig. 3 the energy
density is shown as a function of time at the mid point $z=L_z/2$,
also for the $q=4$ case. This is proportional to the signal that
should be measured by a detector placed at that point.

\begin{figure}[t]
\begin{center}
\includegraphics[width=0.4\textwidth]{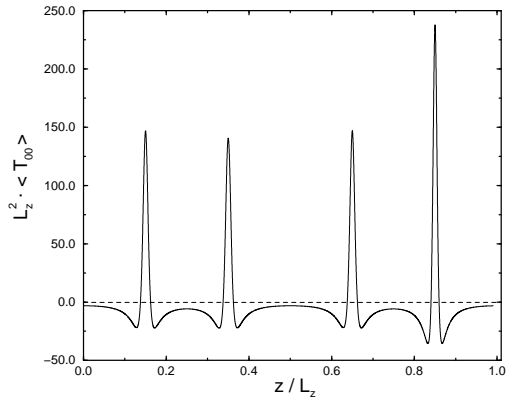}
\caption{One-dimensional energy density profile $\langle T_{00}(z,t)\rangle$ for fixed
time $t/L_z=20.4$ for the $q=4$ case. The amplitude coefficient is $\epsilon=0.01$.}
\label{figure2}
\end{center}
\end{figure}

\begin{figure}[t]
\includegraphics[width=7cm]{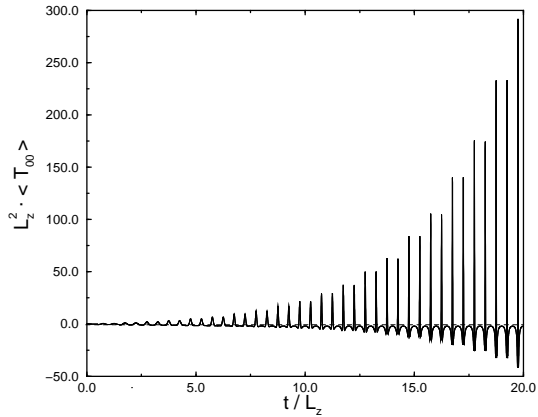}
\caption{One-dimensional energy density profile $\langle T_{00}(z,t)\rangle$ for the
midpoint $z/L_z=0.5$ between the caps of the cavity. The parameters are $q=4$ and
$\epsilon=0.01$.}
\label{figure3}
\end{figure}


\section{Conclusions}

In this paper we have analyzed the nonstationary Casimir effect for
cylindrical cavities with arbitrary section and time dependent
length. Using Hertz potentials, we have shown that for hollow
cylinders the full electromagnetic problem can be treated by
considering two scalar fields, one satisfying Dirichlet boundary
conditions and the other generalized Neumann boundary conditions
on the caps of the cavity. We have derived explicit formulas for
the number of TE and TM photons created during the motion of the
mirror in terms of the Bogoliubov transformation connecting the IN and
OUT basis of the scalar fields. We have also shown the equivalence
between the Hertz potential approach and the dual vector
potentials approach used in previous papers.

Using the TE and TM scalar fields, we rederived results for the
number of photons created in resonant situations for cavities of
rectangular section, and we computed the TE and TM photon creation
in cavities with circular section.

We also considered non-simply connected cylindrical cavities. In
this case, it is necessary to introduce a third scalar field
satisfying Dirichlet boundary conditions to take into account the
TEM waves. We have shown that the dynamics of this field is
equivalent to that of a scalar field in $1+1$ dimensions.
Therefore, all results derived previously in "toy models" are
useful to describe the creation of TEM photons, i.e., TEM modes
provide a realistic realization of the models in $1+1$ dimensions.
This is interesting not only from a theoretical point of view.
Indeed, TEM modes have in general a lower fundamental frequency
than TE and TM modes (this is the case for example for a
rectangular cavity containing an inner cylinder along the longest
direction). This is important since one of the main difficulties
to measure the dynamical Casimir effect is to produce oscillations
of the mirror at twice the lowest frequency of the cavity.
Moreover, as the spectrum of TEM modes is equidistant, the modes
are coupled and the energy density in the cavity develops a very
particular structure  that might be detected more easily than
photons of a given frequency. Taking into account these results,
it may be of interest to excite TEM photons by changing the
effective length of a cavity by irradiating a semiconductor placed
inside it . This could be done by inserting a metallic cylinder
into the cavities used in the experiments of Ref \cite{carugno}.

\section{Acknowledgments}
F. C. L. and F. D. M. were supported by Universidad de Buenos Aires, CONICET,
Fundaci\'on Antorchas and Agencia Nacional de Promoci\'on
Cient\'\i fica y Tecnol\'ogica, Argentina. M.C. and D.A.R.D. thank Carlos Villarreal
for useful conversations on Hertz potentials. We would like to thank the organizers
of this topical issue, Gabriel Barton, Victor V. Dodonov and Vladimir I. Man'ko, for the
invitation to submit a paper.

\end{document}